# Climate change adaption in Chinese ancient architecture

Siyang Li[1,#], Ke Ding[2,4,#], Aijun Ding[2,4,*], Lejun He[1], Xin Huang[2,4], Quansheng Ge[3,4,*], and Congbin Fu[2,4]

[1]Shool of History, Nanjing University, Nanjing, 210023, China

[2]School of Atmospheric Sciences, Nanjing University, Nanjing, 210023, China

[3]Institute of Geographic Sciences and Natural Resources Research, Chinese Academy of Science, Beijing, China

[4]Collaborative Innovation Center for Climate Change, Nanjing, 210023, China

[#]Equal contribution

Correspondence to A. Ding (dingaj@nju.edu.cn) or Q. Ge (geqs@igsnrr.ac.cn)

**SUMMARY**

As an important symbol of civilization and culture, architectures originally were built for sheltering human beings from weather disasters and therefore should be affected by climate change, particularly the associated change in the occurrence of extreme weather events. However, although meteorology has been considered as a factor in modern architecture design, it remains unclear whether and how the ancients adapted to climate change from the perspective of architecture design, particularly on a millennium time scale. Here we show that the periodic change and trend of the roof slope of ancient architectures in northern China clearly demonstrate the climate change adaptation over the past thousand years. We show that the snowfall fluctuation caused by the paleo-climate change was an essential driving factor within the roof modification process of ancient Chinese timber architecture. Our study indicates that climate change may act as a much more important factor on human behaviour than we ever thought.

# Introduction

Societal and cultural responses to climate change during the Holocene have been widely addressed by integrating detailed archaeological data and paleoclimatic records[1-3]. In West Asia, North Africa, the Indus Valley, Europe and many other preindustrial civilizations, climate change (e.g., a sudden or prolonged cooling, a short-time aridification or a persistent multi-century drought) has been examined and reported as a key factor leading to social collapses and the demise of cultures[3-15]. Known as one of the earliest civilisations in the world, the Chinese civilisation went through numerous climate changes[16], which lead to responses in agriculture, population, economy[17-20], even contribute to wars, riots, and the rise and fall of dynasties[21-24].

While focusing on the so-called crises, however, scientific researches have almost neglected how and to what extent the people at that time were affected by the climatic variability and in what ways they acclimated. Chinese traditional architecture, as an important part reflecting everyday life and an integral component of ancient Chinese civilization, closely related to its natural environment all the time, due to its building material and unique system of construction[25-28]. Fortunately, this intensive climate-architecture connection gives us a new way to understand more about the comprehensive impact of climate change on ancient civilizations and people's creative intelligence in their climatic adaptational behaviour.

Roofs, playing important roles in sheltering human beings from rainfall, snowfall, and sunlight, are also the most climate sensitive, vulnerable, and exposed part of buildings[29-33]. Therefore, investigate the historical adjustment of the roofs may gain insights into the relationship between architecture design and climate change adaption[29,34]. However, to our knowledge there are rare of studies to investigate such kind of relationship from a millennium time scale. China, one of the oldest countries, has a unique style in the roofs of ancient architectures. In this study, we reconstructed a chronosequence of roof pitch based on ~200 official architectural remains in northern China and applied high-resolution reconstructed paleo-climatic data[35,36] to explore the impacts of climate change on the roof modification of ancient Chinese architecture and the people's acclimating behaviour from the perspective of dwelling. By an integrated study based on the meteorological reanalysis data and architectural records, we discovered that the key factor for driving the millennial-scale modification of roof is the paleo-climate change. This study urges greater scientific and societal attention as we are facing escalated anthropogenic climate change that calls for affordable, reliable, and resilient climate adaptation strategies, particularly in the Global South given their limited access to adequate climate adaptation resources.

**Results**

**Change of roof pitch over the past thousand years**

The sloping and curved roof, which is an immediately outstanding feature of ancient Chinese architecture, represents the social hierarchy, aesthetic customs and closely related to the beam structure[27]. While retaining its principal characteristics from prehistoric times to the present, the roof modifies and reveals a distinctive appearance of an era (Figure.1A, B). Dating from the mid-late Tang Dynasty to late Qing Dynasty (750 AD to 1750 AD), the official architectural remains are mainly located in the central and eastern regions of northern China (Figure.1C), where belongs to the central area of ancient Chinese culture, in the meantime a geographically semi-humid area of the high climate sensitivity[37-39]. In other words, this unique location controlled by the winter and summer monsoon is affected by both the cold-dry air from the northwest and the warm-wet air from the southeast. Therefore, it tends to suffer from the amplified influence of climatic fluctuation and gives us an excellent chance to explore the impacts of climate change on ancient Chinese architecture.

Surveys since that conducted by Liang (1984) and Chen (1993) have noticed the significant change in roof pitch of ancient Chinese architecture (represent by height span ratio of roof, abbreviated as HSR)[27,40] (Figure. 1A, B and Figure. 2). Many recent studies have further suggested the upward trend of HSR from ~20% in the 8th century to ~35% in the 18th century. It is also reported a significant reduction of the roof curvature around 1600 AD[27,41], along with several analyses of the causes that pointed to the aesthetic changes in Yuan Dynasty (1271-1368AD) and the technical improvement in late Ming and early Qing dynasties (around 1650AD)[27]. However, as can be seen from the reconstructed roof pitch chronosequence (Figure. 2), there exists a more complicated and nonlinear HSR variation with a century-scale fluctuation. During 1100-1200AD and 1300-1650AD, there was an obvious increase of roof pitch (especially the minimum HSR). In contrast, it became relatively unrestricted and went down by the large during 1200-1300 AD, which led to a smoother roof trend around 1300 AD. Consequently, a more reasonable explanation addressing the causes of the detailed, cycle lasting HSR fluctuation should be included in the research of ancient Chinese architecture. In consideration of the similar time scale of roof pitch fluctuation comparing with climate change[42,43] and the high climate sensitivity of ancient Chinese architecture, we applied the HSR data with the reconstructed temperature in central-eastern China by Ge et al.[44,45] to investigate their relationship.

As shown in Figure. 2, the HSR of roof (especially the minimum HSR) has a negative correlation with the reconstructed average temperature in winter half year, along with a nearly 30-year delay (Figure. S1). In cold years, the roofs became steeper (1100-1200 AD, 1300-1750 AD corresponding to the Little Ice Age[46]), while in warm years, the roof pitch descended significantly (1200-1300AD corresponding to the Medieval Warm Period in Europe[46]). Previous studies have shown a 30-year lag between climate change and its responses in human society[10]. Thus, allowing for this reasonable delay, the negative correlation between roof pitch and the reconstructed temperature may suggest an influence of climate change on the design of ancient Chinese architecture.

This "Cold-steeper, warm-smoother" correspondence is not only reflected in the long-term data but also individual cases. A notable example here is the Longmen Temple in Pingshun, Shanxi Province, which is called Museum of ancient Chinese architecture in the deep mountains (Figure. 3, Figure. S2B). It is a rare case for four out of its five major buildings have a relatively fixed, meanwhile unique construction/re-construction time, which belongs to different warm and cold periods during 925 AD to 1504 AD, and well represent the characteristics of HSR trend in different ages (black stars in Figure. 2). According to the historical document of building standards in the Song Dynasty (Yingzao Fashi, 1103 AD)[47], these four buildings in Longmen Temple are supposed to have a basically identical roof pitch around 27%. However, in fact, the two in cold (or quickly becoming cold) years are found steeper (1098AD, 29.67% and 1498-1504AD, 30.50%), while the two in warm (or becoming warm) period smoother (925AD, 23.3% and 1271-1294AD, 26.92%). The case study here illustrates that the "Cold-steeper, warm-smoother" is a widespread phenomenon and the roof pitch responded efficiently to the demand of environmental pressure despite that harmony and coordination is essential in ancient Chinese architecture.

**Driving factors of roof pitch changes associated with climate change**

In the presence of the negative relationship between air temperature and roof HSR, we are particularly interested in uncovering the role of climate change in such a relationship. It is well known that roofs have the primary function of sheltering from sunshine, wind, and precipitation, including snow in winter. First, as enclosed buildings, no evidences show that the roof pitch influences indoor air temperature. Second, considering that these buildings locate in the semi-humid area in northern China, a region with relatively weak annual rainfall, snow load turns into a significant aspect in roof designing for the close relation between roof pitch and the snow removal capacity revealed in either American, Canadian, European, or Chinese building standards[48-52]. Existing literature or news also reported the damage and collapse of ancient buildings by heavy snow, hence cause economic losses and casualties[53]. Therefore, we could make a hypothesis that the relationship between roof pitch and air temperature might reflect the influence from change of snowfall intensity.

To explore the possible relationship between air temperature and snowfall, we analyse the modern reanalysis data during 1981-2019. Since these buildings locate in a region with complex topography, where air temperature and snow should strongly depend on the topographic height[54], such kind of relationship between air temperature and snow could also stand for ancient period considering that the topography hasn't changed a lot in a millennium scale. We select the period of 1110-1210 AD, when the climate turns significantly cold owing to the weakening of solar radiation and the volcanic eruption[55,56], to examine if the roof pitch correlated with snowfall (see Experimental Procedures). We found that the roof pitch of remains (same in the cold period and geographical close, differ in topographic position) is positively related to the local snowfall intensity (Figure. S2A, Figure. S3A). With the increase of snowfall from about 0.2 to 0.35 mm day$^{-1}$, the HSR of the roof goes up from ~25% to ~35%. It thus indicates the significant driving force of snowfall for creating the diversity of the roof pitch. Meanwhile, modern architectural research proves a stronger snow removal capability when the HSR is higher than 20% in China[52], which is also consistent with our analysis in ancient Chinese architecture and technically supports the conjecture of the roof pitch-snowfall intensity relationship. Furthermore, it is worth noting that the variation range of the roof HSR from ~25% to ~35% is almost the same as it ranged from mid-late Tang Dynasty to late Qing Dynasty (750 AD to 1750 AD) (Figure. 2), which further supports that the change of snowfall caused by climatic fluctuation may be the essential factor driving the modification of roof pitch over the past thousand years.

Prior to statistically assess this correspondence between climate change and the roof pitch modification, we developed a proxy of estimated snowfall intensity anomaly for the absence of the reconstructed historical snowfall data in northern China. Based on the reconstructed average air temperature anomaly and a deduced quantitative relation of -0.074 mm day$^{-1}$ K$^{-1}$ between the snowfall and air temperature (utilizing the reanalysis data from 1989-2019, shown in Figure. S3B), we roughly estimated the snowfall intensity anomaly from 950 AD to 1750 AD (see Experimental Procedures).

This estimation, as shown in Figure. 4, agrees well with the qualitative standard of snowfall anomaly developed by Chu et al. (2008) upon the statistics of snowfall events in historical documents[36]. It is evident that around 1300AD in the warm period, there is almost no such record, whereas in 1300-1750AD during the Little Ice Age, the continuous descendent in air temperature is found accompanied with a sharp increase of recorded snowfall events[57].

Furthermore, through compositing the statistically estimated snowfall anomaly and the quantitative relation between snowfall intensity and roof pitch, we further estimated the snow-driven HSR from 950 to 1750 AD (see Experimental Procedures). What stands out, as presented in Figure. 4, is the considerably close correspondence between the estimated roof HSR and the minimum roof HSR deriving from building remains, as well as the estimated HSR (no smooth) exactly reproducing the low HSR values around 1250 AD, 1430 AD, 1540 AD and the high HSR values around 1180 AD, 1490 AD, 1650 AD. The positive correlation indicates that the fluctuation of snowfall intensity owing to climate change being an essential driving factor in the modification of roof pitch over the past thousand years. During warm periods, the roof pitch tended to be more unrestricted due to the less snowfall intensity and accordingly showed a wider range and a lower tendency in roof HSR, while when the cold period came, more massive snowfall asked for a steeper roof in newly-built houses to remove the snowfall more efficiently. This conclusion could be further confirmed by the reduction of roof curvature occurred around 1600 AD (corresponding to Little Ice Age, and the technique for building roofs improved simultaneously) [27, 41], for a less curvature of roof also enhances the snow removal capacity. From another perspective, the statistical roof pitch gives us a new method to quantify the snowfall changes during cold periods, which is difficult to obtain from numerical models and commonly used meteorological historical data like tree rings, stalactite and so on. As for the relatively large bias between the estimated and measured minimum HSR during 1550 AD to 1750 AD, we assume the difference comes from the underestimate of extreme snowfall intensity[58,59,60], and the smooth of HSR.

It could be argued that why the ancient people did not rebuild their roofs to adapt to the increased snowfall intensity when cold periods came, if the snowfall anomaly being the main factor that led to the roof pitch modification over the past thousand years. This contradictory situation could be attributed to the economic cost in roof construction, which attributes to even up to half of the total labour and materials cost of an ancient Chinese building[47]. Therefore, repairing became a regular choice for a built one as the roof was damaged or deformed, while designing and constructing a steeper roof for a new one in cold years characterised the fluctuation process of roof pitch which we have noticed today. Moreover, one may wonder why the ancient Chinese people didn't just keep a steeper roof to avoid the possible damage even in warm years. In addition to considering the aspects of cost and a diverse need of sunshine and rainfall sheltering, another possible explanation may come from the study of the collective memory of natural disaster, which suggests that people could remember a catastrophe for one generation to two. Then the knowledge of the event fades away from memory while the people behave the way they did just before the disaster took place[61]. Also, this time span of memory and knowledge explains the delay of modification of roof HSR in Figure. 4, compared to the fluctuation of temperature when the cold period came. Therefore, through the above analysis, we point out that the abnormal snowfall caused by climate change is the main factor that drives the modification of roof pitch within the development of ancient Chinese architecture over the past thousands of years.

## Discussion

The sloping and curved roof, one of the most prominent features of ancient Chinese timber architecture, has previously been reported an overall upward trend in roof pitch and a significant reduction of curvature around 1600AD. In recent studies, the modifications in roof, which formed different appearances of architecture within different periods, have been generally attributed to the varies aesthetic customs or the improvement of building technology. This research, however, to our knowledge, for the first time identified a non-monotonic increase of the roof HSR from ~200 remains over millennia. Our integrated study quantitatively revealed that the main driving factor in the roof pitch and curvature modification process is the fluctuation of snowfall intensity caused by climate change over the past thousand years.

In addition to giving a reasonable account of the driving force within the roof modification process of ancient Chinese architecture, this study has also sought for a more comprehensive understanding of the influences of climate change. Rather than identifying the relationships between abrupt climate events and social crises or cultural collapses, this study focuses more on the ancient people and the aspects concerning their everyday life. The results of our investigation end up confirming the responding performance of roof modification to the climatic fluctuations for more than one thousand years. Thus, indicating an intelligent long-term adaptive behaviour of ancient people to adjust their buildings for a more stable and suitable roof formation when faced with the various weather conditions caused by climate change.

Moreover, it therefore can be assumed that the significant aesthetic changes and technological improvements showed in buildings between different eras were also consequently following the driving demands according to climate change, for the essential aspects that related to like the construction safety, durability, and cost. Therefore, this study suggests that the climate change is an essential background, even foundation for a wider extent of research, including the ancient people and their material and even intangible culture like aesthetics and technology.

The wisdom of climate change adaption in architecture design discussed in this study came at the expense of economic losses and casualties. Today, we are facing the challenge of a more intensive climate change[62], but we have been improving our capability in predicting and adapting the climate change actively. This study suggested that the designs for buildings or even the cities should consider more about the adaption to the comprehensive environment change in the future[63]. Apart from the immediate heating impact of global warming, the occurrence of large number of extreme weather events, including heavy snow, drought, flood, and other natural disasters, is another severe problem that we are facing[64-67]. To reduce the losses caused by extreme weather events, designers should consider factors of local climate and natural disasters carefully according to climate prediction[29,68-70]. Furthermore, due to the intensity of extremes are always underestimated by nowadays models[71,72,73], the urban and architectural design codes should be stricter and more comprehensive to cope with the increasing extreme events. Especially for some of developing countries, e.g., in the Global South, where wood is widely used as the primary building material, the government should pay more attention to the adaptive strategies on building designs, as timber architecture is vulnerable and more sensitive to the climate change. Reinforcing or rebuilding with new materials, that are more economical and adaptive to the harsh natural environment, may be a feasible way in dealing with the dramatic impacts of climate change[74,75,76].

## EXPERIMENTAL PROCEDURES

### Statistics of roof height span ratio

To static the HSR in the history and reduce the noise due to the relatively less data, we bin the HSR data on the south region with a 30-year resolution and smooth the 5%, 95% percentages and mean value in each bin by using a locally weighted regression method. The reason for choosing the data on the south region is the remains mainly leftover in the south region, and the climate of the two regions is quite different, which can be seen from the precipitation amount (Figure. 1C).

### Relations between HSR and snowfall

To estimate the relations between HSR and snowfall, in this paper, we analyse the HSR during 1110 AD-1210 AD when it is freezing. One reason is that in the cold period, the roof pitch will depend more on snowfall. The other reason is that there are enough remains left in regions near but with different snowfall in that period, making the noise of time and locations could be neglected. The snowfall is higher in the windward slope of the mountains and lower in the valleys is mainly due to the topography[54], making the snow distribution will not change a lot in the history. Therefore, by using the snowfall data from reanalysis dada (ERA5-Land), the comparison between the roof's HSR on the mountains and valleys could quantify the snow load of the roof (Figure. S3A). And because the snow load of the roof is almost linearly related to roof slope [48], the relation between HSR and snowfall will not have a problem, even though the snowfall will have some fluctuations. Note: ERA5-Land is a reanalysis dataset providing a consistent view of the evolution of land variables over several decades with a resolution of 0.1 degree. And the unite for snowfall the is height of water equivalent, which means the depth the water would have if the snow melted and was spread evenly over the grid box.

**Reconstruct the data of snowfall anomaly and HSR**

To estimate the snowfall anomaly during 950 AD to 1750 AD, we use the reconstructed temperature anomaly data and the relationship between snowfall and temperature during 1981-2019. The equation is:

$$\text{Snowfall\_anomaly}_{estimate} = R_{snow} * T_{anomaly} \quad (1),$$

where $R_{snow}$ represents the growth rate of snowfall with temperature in winter half-year, calculated by using reduced major axis regression for temperature and snowfall value at locations of remains in the south region during 1981-2019 (Figure. S3B). The temperature and snowfall data are from ERA5-Land. The $T_{anomaly}$ represents the reconstructed temperature [35].

To estimate the HSR in history, we use the estimated snowfall anomaly and the relations between HSR and snowfall. The equation is:

$$\text{HSR}_{estimated} = \text{HSR}_0 + R_{HSR} * R_{snow} * (T_{anomaly} + 30) \quad (2),$$

where $\text{HSR}_0$ represents the HSR baseline and calculated as the mean value of HSR during 950 AD -1000 AD. $R_{HSR}$ represents the growth rate of HSR with snowfall. $R_{HSR}$

is calculated by using reduced major axis regression for snowfall and HSR value during 1110-1210 AD (Figure. S3A). And we delay the estimated HSR by 30 years due to the 30-year lag between climate change and its responses in human society[10].

**Data availability**

The reconstructed temperature anomaly data is from Ge[25], the historical snowfall events data is from Chu[26]. The ERA5-Land data could be download from the website: https://cds.climate.copernicus.eu/cdsapp#!/home. The HSR data is from papers which are listed in Supplementary Table 1.

**Acknowledges**

This study was funded by the Collaborative Innovation Center for Climate Change. We are grateful to the group for ERA5_land data and the group of Guoqiang Chu. We also thank Chi Zhang and Rui Chen for their help.

**Author contributions**

Siyang Li, Ke Ding, and Aijun Ding conceived the overall idea. Lejun He contributed to data collection. Xin Huang help to improve pictures. Quansheng Ge guided the data analysis and discussion on its relationship with the past climate change. Siyang Li and Ke Ding made most of the analysis and wrote the manuscript with contributions from all authors.

**Competing interests**

The authors have no competing interests.

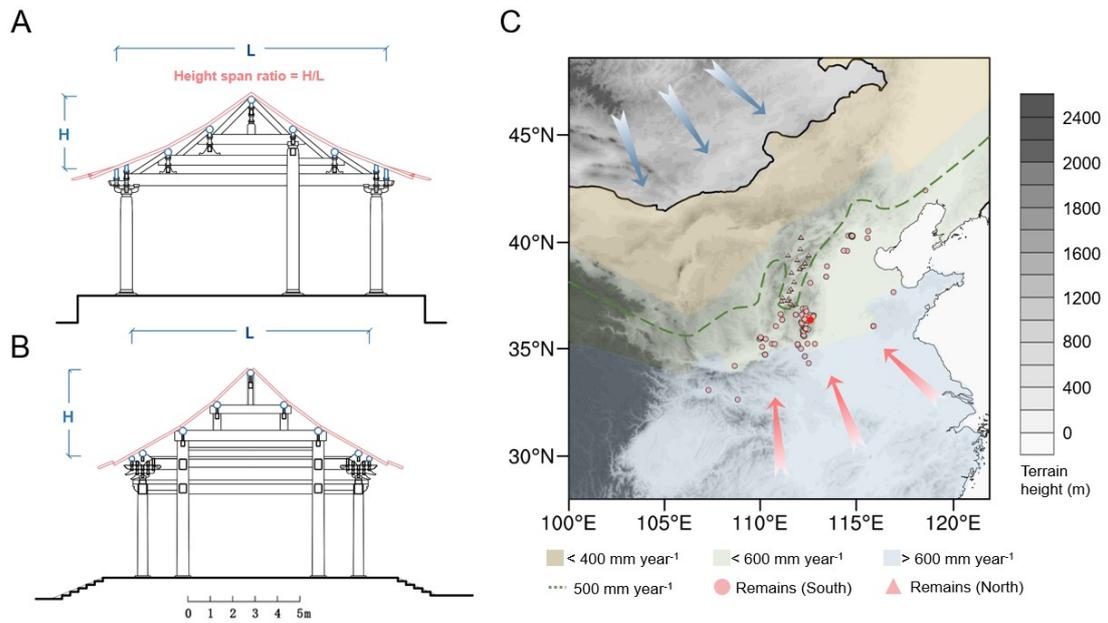

**Figure. 1 The typical Chinese timber architecture and the locations of the remains.**
**A**, The typical structure of a Chinese timber architecture at Song Dynasty (around 1103 AD) and the definition of height span ratio (HSR). **B**, Same as (**A**) but at Qing Dynasty (around 1700 AD). **C**, the locations of remains during 750-1750 AD. The light red circle markers show the remains on the south region (precipitation amount larger than 500 mm year$^{-1}$) while the light red triangle markers on the north region (precipitation amount less than 500 mm year$^{-1}$). The dark red marker gives the location of Longmen temple. Note: The black and white color represents terrain height and the colorful shading areas represent the locations with different precipitation amount. The green dashed line shows 500 mm isoline of precipitation. The blue and red vectors mean the cold-dry and warm-wet air from the north and the south.

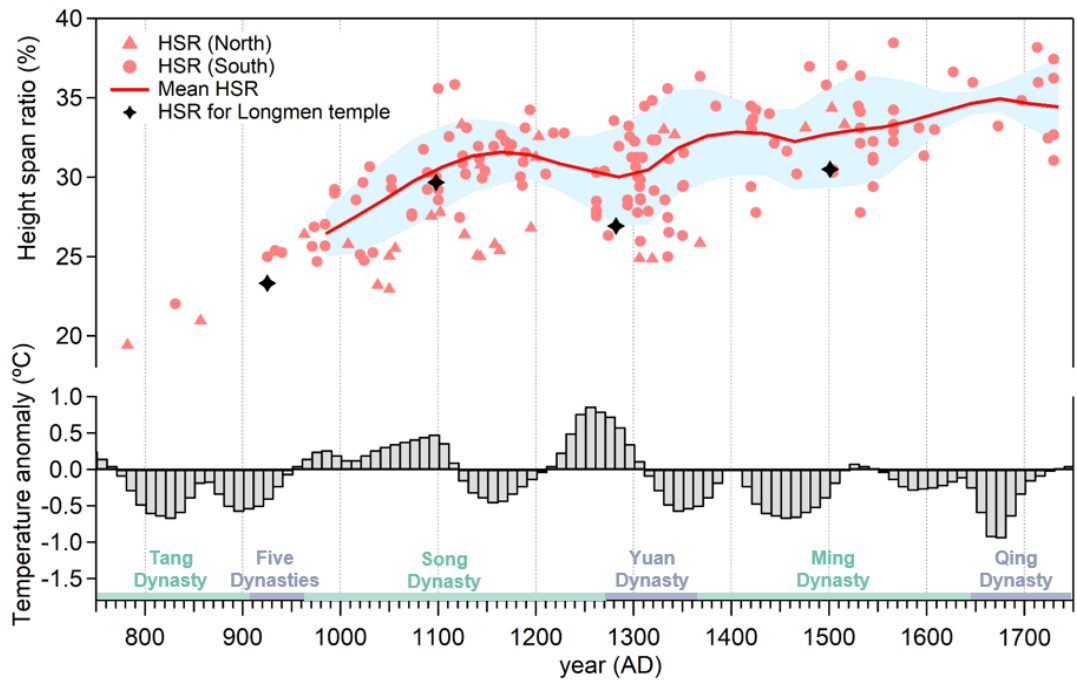

**Figure. 2 Time series of HSR and air temperature anomaly during 800-1750 AD.** The triangle and circle markers show the HSR of the remains in the north and south region (locations definitions are shown in Figure 1). The blue shaded area marks the 5th-95th percentiles range of HSR which is smoothed using a locally weighted regression. The red line shows the mean HSR in south region. The black stars give the HSR in Longmen temple. And the gray bars show the temperature anomaly in winter-half year in central-eastern China.

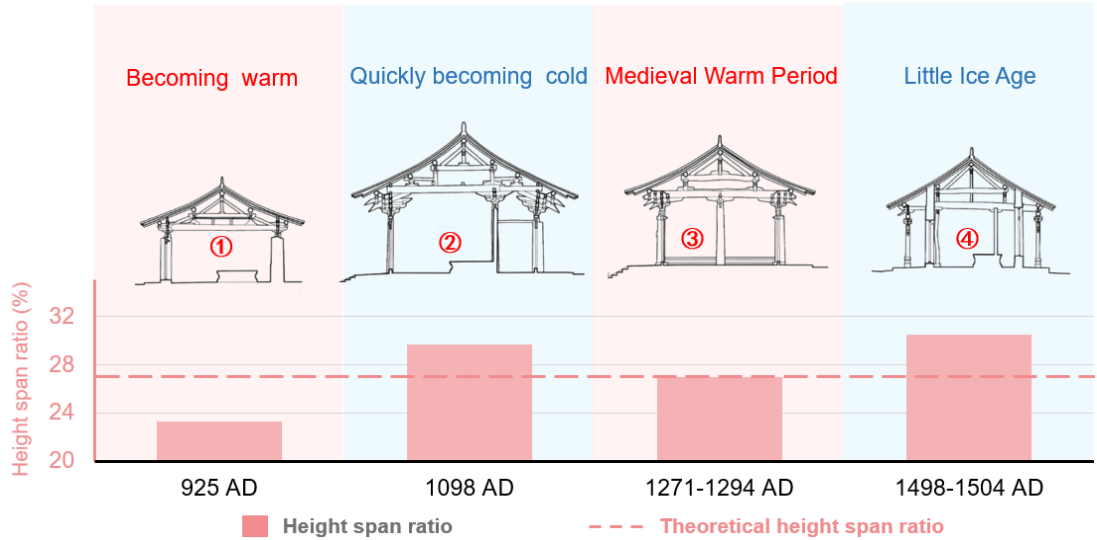

**Figure. 3 Variations of HSR during 925-1504 AD in the Longmen temple.** The building designs show the four buildings belong to different warm and cold periods in Longmen temple in chronological order in upper panel. The red bars give the corresponding HSR, and the red dashed line gives the theoretically HSR. The HSR of different buildings are also shown in Figure. 2. The location of Longmen temple and the layout of buildings are shown in Figure. 1 and Figure. S2.

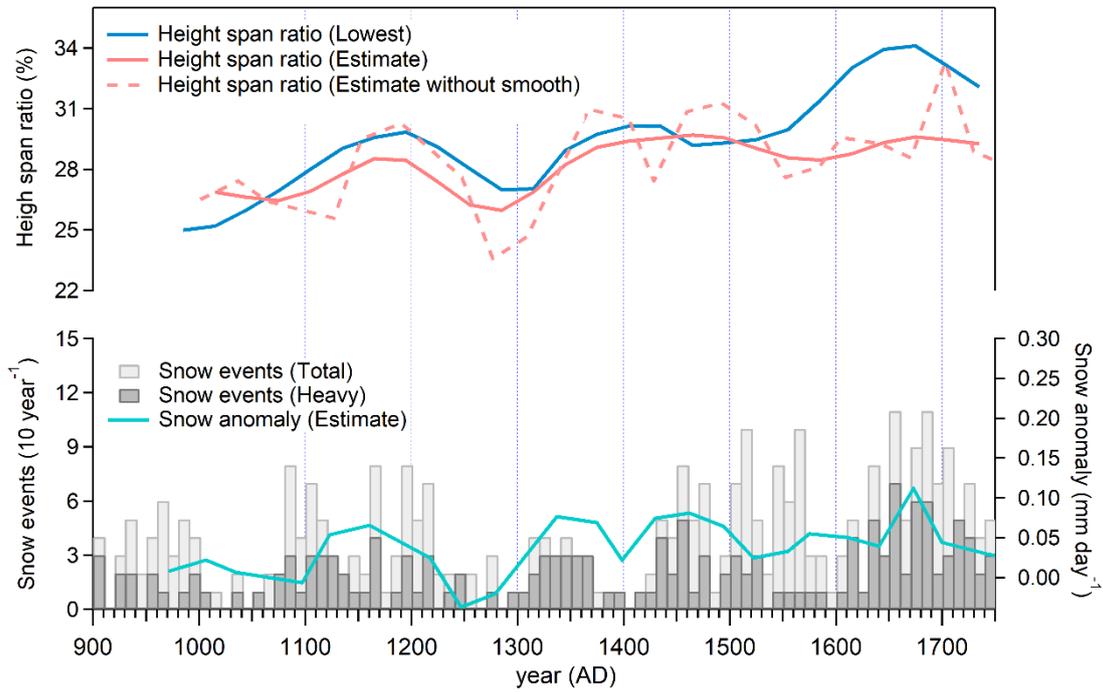

**Figure 4 Comparisons of snowfall and HSR between estimated data and observations.** The solid blue and red lines show the smoothed observed minimum HSR (also shown as the lower bound of the blue shaded area in Figure. 2) and smoothed estimated HSR, while the dashed red line gives the estimated HSR without smooth. The light and dark gray bars give the number of total and heavy snow events in historical records in central-eastern China, while the green line shows the estimated snow anomaly. Note: The unite for snow anomaly is height of water equivalent, which means the depth the water would have if the snow melted and was spread evenly over the grid box.

# Supplemental Figures

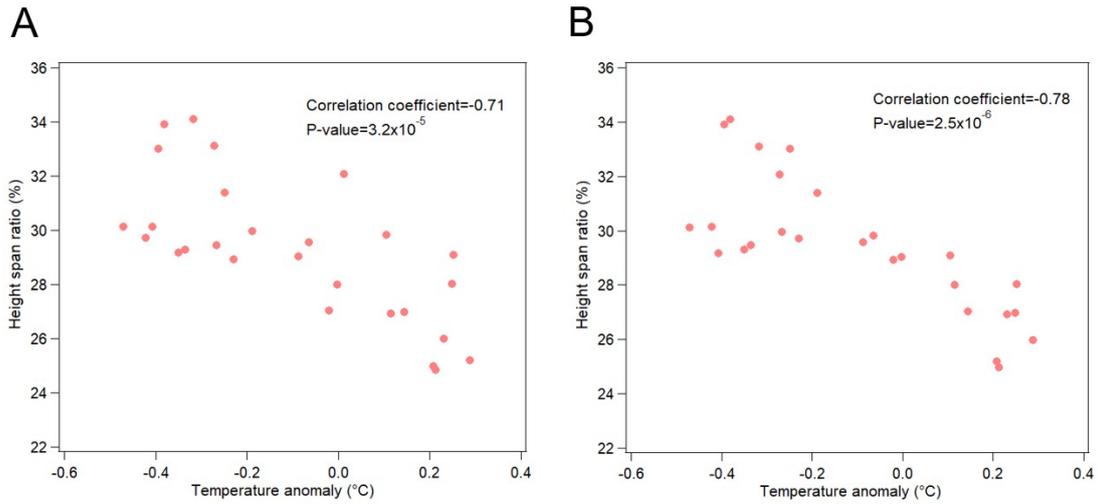

**Figure. S1 Correlation between temperature anomaly and HSR. A,** The correlation between the smoothed temperature anomaly and the lowest HSR in history. **B**, Similar with (**A**) but for 30-year delay correlation. Note: The data for calculation correlation coefficient is smoothed by a locally weighted regression method.

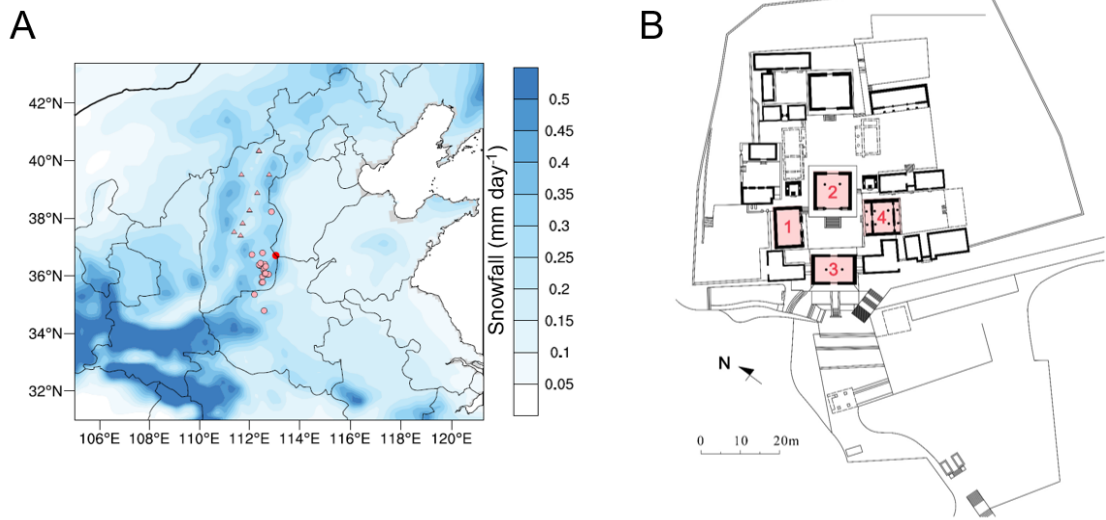

**Figure. S2 Distribution of snowfall and the layout of Longmen temple. A**, The averaged snowfall in winter during 1981-2019 with the light red triangle and circle markers show the locations of remains used to quantify the relationship between HSR and snowfall in Figure. S3A. **B**, The layout of Longmen temple, with the numbers indicate the buildings in Figure. 3. The location of Longmen temple is shown in (**A**) with a dark red circle marker.

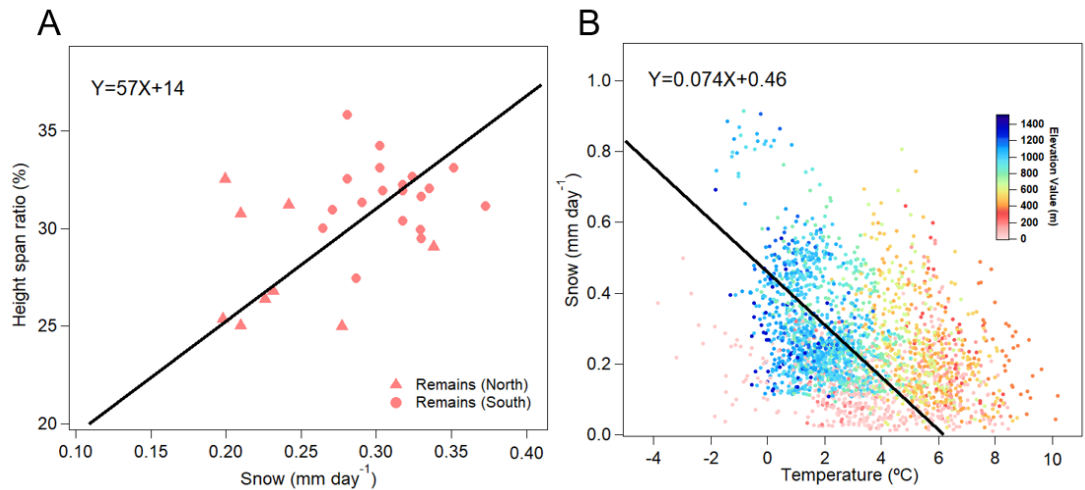

**Figure. S3 Relationship between HSR, snowfall, and air temperature.** Scatter plots of (**A**) Averaged snowfall (during 1981-2019) versus HSR of remains in cold period (1110 AD -1210 AD) and (**B**) temperature and snowfall in winter half year at the locations of remains in the south region (defined in Figure. 1C) during 1981-2019. The snowfall and temperature data are from ERA5-Land dataset. The black lines give the reduced major axis (RMA) regression for the data sets in each panel with the fitting function labelled on the left upper corner.